\documentclass{jetpl}
\usepackage[cp1251]{inputenc}
\usepackage[russian]{babel}
\usepackage{bm}
\usepackage{amsmath}
\usepackage{amssymb}
\usepackage{amscd}
\usepackage[dvips]{graphicx}
\usepackage{epsfig}

\twocolumn

\title{Finite-size effect in shot noise in hopping conduction}

\rtitle{Finite-size effect in shot noise
\ldots}
\sodtitle{Finite-size effect in shot noise in hopping conduction}

\author{E.S.~Tikhonov$^{1,2}$, V.S.~Khrapai$^{1,2}$, dick@issp.ac.ru, D.V.~Shovkun$^{1}$, D.~Schuh$^{3}$}

\rauthor{E.S.~Tikhonov, V.S.~Khrapai, D.V.~Shovkun, D.~Schuh}
\sodauthor{E.S.~Tikhonov, V.S.~Khrapai, D.V.~Shovkun, D.~Schuh}
\address{1 -- Institute of Solid State Physics, Russian Academy of
Sciences, 142432 Chernogolovka, Russian Federation}
\address{2 -- Moscow Institute of Physics and Technology, Dolgoprudny, 141700 Russian Federation}
\address{3 -- Institute of Applied and Experimental Physics, University of Regensburg, D-93040 Regensburg, Germany}

\abstract{We study a current shot noise in a macroscopic insulator based on a two-dimensional electron system in GaAs in a variable range hopping (VRH) regime. At low temperature and in a sufficiently depleted sample a shot noise close to a full Poissonian  value is measured. This suggests an observation of a finite-size effect in shot noise in the VRH conduction and demonstrates a possibility of accurate quasiparticle charge measurements in the insulating regime.}

\begin{document}

\maketitle

As first shown by Schottky for the case of a vacuum tube, electric current can be viewed as a sequence of uncorrelated pulses corresponding to arrivals of individual electrons at the anode~\cite{blanter}. A mean-squared current fluctuation (shot noise) in this random Poissonian process has a spectral density of $S_I=2qI$, where $q\equiv e$ is the elementary charge and $I$ is the average current. A direct shot noise measurement of the charge $q$ of a quasiparticle is intriguing in application to various solid state materials, where $q$ can be renormalized by interactions ($q\neq e$)~\cite{reznikov,glattli}. This list includes nontrivial many-body insulating states in charge-density wave compounds~\cite{monceau}, in cooper pair insulators~\cite{shahar,baturina} and in the bulk of a two-dimensional (2D) system in fractional quantum Hall effect~\cite{5/2}.

The above concept of the charge measurement in solid state can be complicated by a non-Poissonian statistics of the current flow~\cite{levitovlesovik1993}, as characterized by a Fano factor $F\leq1$ in the expression for the noise spectral density $S_I=2FqI$. At low enough temperatures ($T$) the transport in the band of localized states occurs via a variable range hopping (VRH) conduction~\cite{schklovskii_efros}. Unlike the case of coherent transport at $T=0$~\cite{beenakerreview}, the Fano factor in the VRH regime is not universal, which is a fundamental problem for the charge measurements.

In long enough samples, the experiments in 2D VRH regime~\cite{kuznetsov,camino} find that the Fano factor decays with the sample length roughly as $F\propto L^{-1}$. This behavior is qualitatively explained by the averaging of the Poissonian noises associated with different hops, analogous to noise averaging in a one-dimensional array of $N$ identical tunnel junctions $F=1/N$~\cite{korotkov_PRB2000}. In the VRH conduction regime, however, the hopping rates are spread exponentially wide, so that only the most resistive  hops (so called hard hops) are important for the noise averaging~\cite{kuznetsov,camino}. A typical distance between these hard hops, known as a correlation length $L_C$ of the critical cluster~\cite{schklovskii_efros}, represents a length scale for the Fano-factor~\cite{kuznetsov,camino} in long ($L\gg L_C$) samples:  $F\approx(L/L_C)^{-1}$.  Numerical calculations support this qualitative picture~\cite{sverdlov,korotkov_JPhysCM1999}. The shot noise in the opposite limit $L\leq L_C$ is not understood. On one hand, the numerical results~\cite{korotkov_JPhysCM1999} suggest that in short samples the Fano factor remains sub-Poissonian ($F\approx0.7$). On the other hand, the experiments do not exclude the full Poissonian noise value in sub-micrometer sized samples~\cite{kuznetsov,savchenko_PSS2004}.

Here we investigate the shot noise in an insulating state of a macroscopic 2D electron system of a GaAs/AlGaAs heterostructure. There is no doubt that $q\equiv e$ in this system, which makes it an ideal test bed to study the current statistics in the insulating state. The shot noise close to the Poissonian  value is achieved in the VRH conduction regime at low $T$ and in a sufficiently depleted sample. We interpret this as a manifestation of  a finite-size effect in shot noise at $L\sim L_C$ and support by transport measurements. Available theories~\cite{sverdlov,korotkov_JPhysCM1999,korotkov_PRB2000} have difficulties to explain these observations and an alternative simplified explanation in classical terms is proposed. Our results open up a possibility of accurate quasiparticle charge measurements in nontrivial many-body insulating states~\cite{monceau,shahar,baturina,5/2}.

\begin{figure}[t]
 \begin{center}
  \includegraphics[width=0.8\columnwidth]{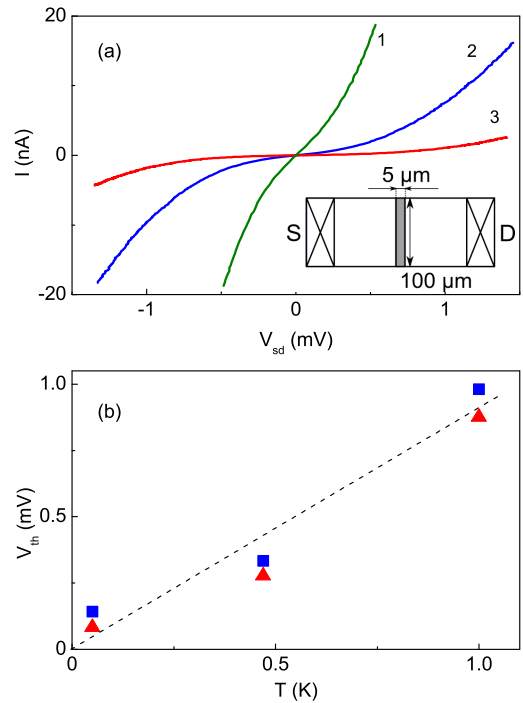}
   \end{center}
  \caption{Fig.~\ref{fig1}.~(a): $I{\text -}V$ measurements. $I{\text -}V$ curves at $T=0.56$~K for a set of gate voltages in sample 1:  $V_g=-0.302$~V (1), $V_g=-0.308$~V (2), $V_g=-0.339$~V (3). Inset: the sketch of the sample layout used. (b): Nonlinearity vs temperature.  The $T$ dependence of the threshold bias voltage where the $I{\text -}V$ curves deviate from the linearity by 20\%.  The triangles and squares correspond, respectively, to $R_\Box=1.6$~M$\Omega$ and $R_\Box=18$~M$\Omega$ (at $T=0.47$~K).}\label{fig1}
\end{figure}

Our samples are based on a two-dimensional electron system (2DES) in GaAs/AlGaAs buried 34 nm below the surface. The as-grown electron density and mobility (at 4.2~K) of the 2DES are, respectively, $3.5\times10^{11}$~cm$^{-2}$ and $2.8\times10^{5}$~cm$^2$/Vs. A metallic front gate is used to define an insulating strip in the 2D channel with the length of $L=5$~$\mu$m along the current flow and the width of 100~$\mu$m (see the inset in Fig.~\ref{fig1}). We measured two nominally identical samples and obtained basically the same results reproducible in respect to a thermal recycling.

A two-terminal resistance and $I{\text -}V$ curves were measured with a low-noise 100 M$\rm\Omega$ input resistance preamplifier and a $\sim300$~$\Omega$ series resistance was subtracted. For noise measurements the sample was connected in series with a load resistor $R_0=$1~k$\Omega$. A voltage noise on $R_0$ was amplified by a set of rf-amplifiers and detected in the frequency band 10-20~MHz. A total gain of the circuit was about 70~dB. The first cascade of amplification was represented by a home-made low-$T$ amplifier placed nearby (about 1 cm) the sample. Analyzing the data we treated the load resistor, the ohmic contacts and the 2D channel as independent noise sources. The respective rf-impedances are equal to the dc differential resistances~\cite{blanter}, obtained by a numerical differentiation of the $I{\text -}V$ curves. The absolute calibration was utilized via the Johnson-Nyquist noise measurements, with both the $T$ and the sample resistance varied. In this approach the influence of a shunt capacitance of $\sim$5~pF is automatically absorbed into the load-dependent gain. An example of one such measurement is shown in supplemental material. The noise measurements were performed in a liquid $^3$He cryostat in the range $0.5$~K$\leq T\leq4.2$~K. A $^3$He/$^4$He dilution refrigerator was mainly used to extend the $T$-range of the resistance measurements down to 60~mK.

Care was taken to ensure that the measured shot noise originates from charge discreteness and is not influenced by noises of different origin. With this goal we chose the frequencies at least two orders of magnitude higher than in previous experiments~\cite{kuznetsov,camino,savchenko_PSS2004}. Also the heterostructure used was relatively clean, so that the samples did not exhibit sizeable mesoscopic fluctuations even at low-$T$ and strong depletion. As a result, we were able to completely get rid of the $1/f$ and telegraph-like noises~\cite{kuznetsov_JETPL}, such that $S_I\propto I$ and is frequency independent in the range 10-100~MHz.

In fig.~\ref{fig1}a we plot $I{\text -}V$ curves measured at $T=0.56$~K for a set of gate voltages $V_g$. A minor asymmetry of the $I{\text -}V$ curves in respect to the origin is related to a grounding of the drain contact. As a result the effective gate voltage is more positive at negative currents, so that the negative $I{\text -}V$ branches are more conducting. All the $I{\text -}V$s are strongly nonlinear and the nonlinearities strengthen with the sample depletion. The nonlinearities are more pronounced at lower $T$, that can be characterized by a $T$-dependence of the threshold bias voltage $V_{th}$. In the absence of a clear threshold behavior we define $V_{th}$ as the bias voltage corresponding to a 20\% deviation of the  $I{\text -}V$ from the linear dependence at small $V$. The $V_{th}$ increases approximately linearly with $T$, as shown in fig.~\ref{fig1}b. This observation is insensitive to the criterium used to define $V_{th}$.

\begin{figure}[t]
 \begin{center}
  \includegraphics[width=0.8\columnwidth]{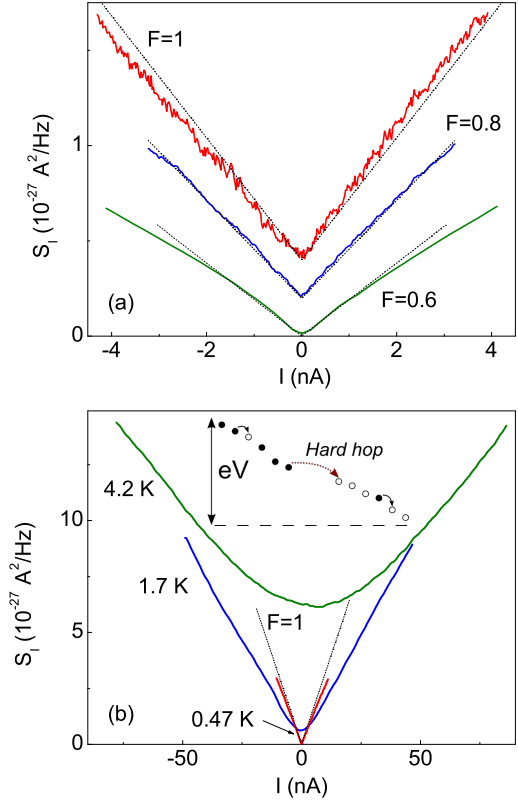}
   \end{center}
  \caption{Fig.~\ref{fig3}.~Shot noise measurements. (a) -- Shot noise spectral density as a function of current at $T=0.56$~K (sample 1). The resistivity (the Mott temperature) from top to bottom are: $R_\Box=58$~M$\Omega$ ($T_0\approx300$~K), $R_\Box=8.8$~M$\Omega$ ($T_0\approx140$~K), $R_\Box=1$~M$\Omega$ ($T_0\approx40$~K). The dashed lines are fits used to extract the Fano factor. The scales on both axes are reduced by a factor of 50 (5) for the lowest (middle) curve and the two upper curves are vertically offset in steps of $2\times10^{-28}{\rm A^2/Hz}$. (b) -- Shot noise spectral density as a function of current for three temperature values and $R_\Box=26$~M$\Omega$ (at $T=0.47$~K), $T_0\approx300$~K (sample 2). The dashed guide line corresponds to $F=1$. Inset -- a sketch of  hopping along a quasi-1D filament of the VRH network with a hard-hop in the middle. The empty/occupied localized stated are depicted as, respectively, the empty/filled circles. }\label{fig3}
\end{figure}

Next we turn to the shot noise measurements in this strongly nonlinear regime. Fig.~\ref{fig3}a shows the noise spectral density as a function of current at $T=0.56$~K for a set of gate voltages. The dependencies $S_I(I)$ are almost symmetric in respect to the current reversal and linear at not too high $|I|$, which is characteristic for the shot noise. The Fano factor determined from this linear region is seen to increase with the sample depletion.
Note that in order to tune $F$ between 0.6 and 1 one has to vary the linear response resistance by about two orders of magnitude. Fig.~\ref{fig3}b demonstrates the $T$-dependence of the shot noise at fixed $V_g$. At higher $T$ the Fano factor is determined outside a crossover region between the equilibrium Johnson-Nyquist noise and the shot noise. Here $F$ increases from $\approx0.4$ to $\approx0.9$  when the sample is cooled down from 4.2~K to 0.47~K. The central result of our paper, shown in fig.~\ref{fig3}, is that at low enough $T$ and for a strong enough depletion the shot noise in a macroscopic sample can reach the maximum possible Poissonian value $F=1.0\pm0.1$. As argued below, this behavior can be interpreted as an observation of a finite-size effect in shot noise
in the VRH conduction. The observation of the Poissonian current statistics in the VRH conduction in a macroscopic sample is remarkable. In particular, this result opens up a possibility for measurements of $q$ in nontrivial many-body insulating states~\cite{monceau,shahar,baturina,5/2} where it might differ from $e$.
\begin{figure}[t]
 \begin{center}
  \includegraphics[width=0.8\columnwidth]{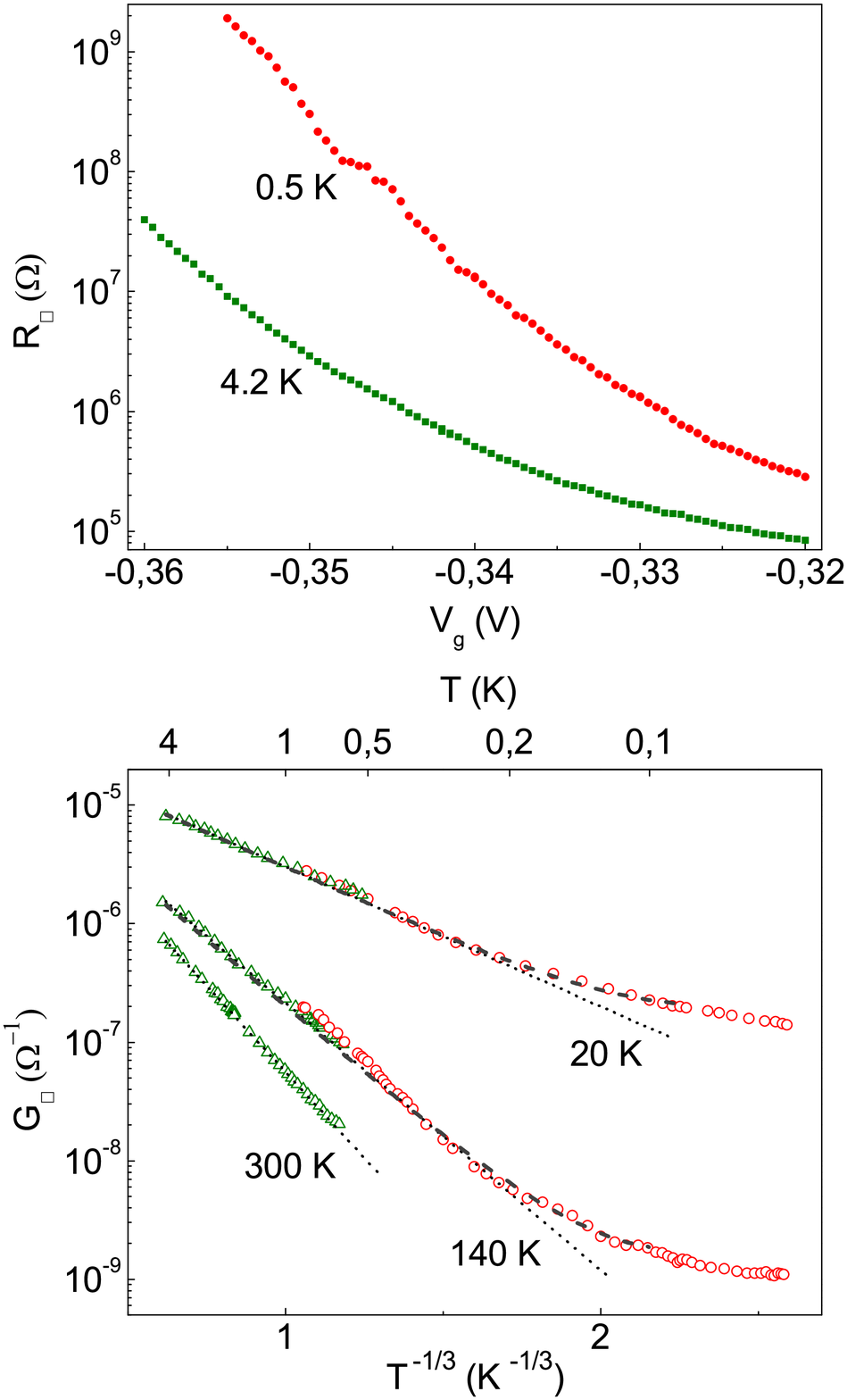}
   \end{center}
  \caption{Fig.~\ref{fig2}.~Linear response resistance/conductance. (a) -- Gate voltage dependencies of the resistivity at two different temperatures (sample 1). (b) -- $T$ dependence of the conductivity in sample 1 for three different depletions (see text). The best fits to the Mott VRH law are plotted as the dotted lines and the respective values of $T_0$ are shown nearby. The fits to the low-$T$ deviations from the Mott law on the two upper curves are shown by the dashed lines. The fit parameter $T_P$ equals 0.14~K/0.2~K  for the top/middle curves (see text). }\label{fig2}
\end{figure}

We would like to stress that in our experiment the shot noise stays close to the Poissonian value even at the strongest depletions~\cite{remarkPoisson}. This is in contrast to a low-frequency random telegraph-like noise in hopping regime~\cite{kuznetsov_JETPL} which causes deviations from the linear dependence $S_I\propto I$~\cite{kuznetsov} and super-Poissonian noise values~\cite{Safonov_PRL}. It is a proper choice of the frequency range  that allowed us to get rid of the spurious noises and study the fundamental properties of the current statistics in the VRH regime.

Gate voltage dependencies of the linear response resistivity $R_\Box$ are plotted in fig.~\ref{fig2}a for two temperatures 4.2~K and 0.5~K. The resistivity is found to strongly increase at sample depletion without significant mesoscopic fluctuations. This data demonstrates a roughly exponential dependence of $R_\Box$ on the carrier density and indicates a strong localization of the electrons. In fig.~\ref{fig2}b we plot the conductivity $G_\Box\equiv1/R_\Box$ as a function of $T$ for different values of $V_g$. With decreasing $T$ the conductivity drops by 1-2 orders of magnitude. Above 0.2~K the $T$-dependencies are best described by the Mott VRH law in 2D: $\ln{G_\Box}\propto-(T_0/T)^{1/3}$ (dotted lines). Here $T_0=13.8/(k_Bga^2)$ is the Mott temperature, $g$ -- the density of states at the Fermi level and $a$ -- the localization radius~\cite{schklovskii_efros}. As seen from fig.~\ref{fig2}b, $T_0$ increases with the sample depletion, which we associate with the decrease of $a$ and $g$. Note that a thermal recycling typically caused some shift of the gate voltage position of the mobility edge. For this reason, the data  from different cryostats were taken at different $V_g$, so that the $T$-dependencies coincide in the range where they overlap. Such data are shown by different symbols in fig.~\ref{fig2}b. At low $T$ we observe deviations from the Mott VRH law and the $T$-dependencies slow down. We have checked via noise measurements that the electronic $T$ follows that of the bath down to $\approx120$~mK (see supplemental material), i.e. the deviation is unlikely to be caused by an electromagnetic pick-up.

\begin{figure}[t]
 \begin{center}
  \includegraphics[width=0.8\columnwidth]{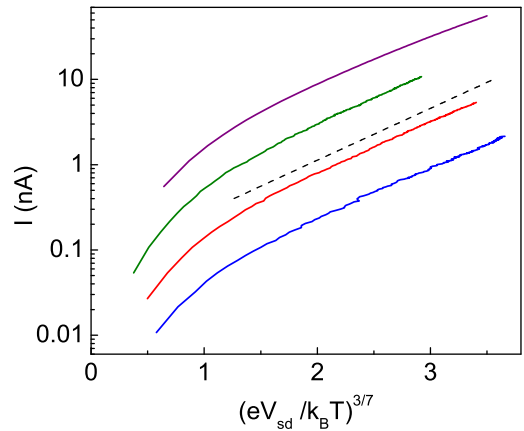}
   \end{center}
  \caption{Fig.~\ref{fig4}.~Nonlinear VRH conduction. Nonlinear $I$-$V$ curves (solid lines) at $T=0.47$~K are well described by the theory of Ref.~\cite{shklovskii_1976}. The curves are obtained for a set of $R_\Box\approx 0.5$, 1.6, 6, 19~M$\Omega$, respectively, from top to bottom. The slope of the dashed line corresponds to    $L_C\approx11$~$\mu$m, see text.}\label{fig4}
\end{figure}

The localization radius $a$ can be evaluated from a measurement of the nonlinear $I$-$V$ curves. In moderate electric fields, the theoretical predictions differ depending whether a typical hop~\cite{hill} or the hard hop~\cite{shklovskii_1976} is assumed to be relevant in the nonlinear regime. Our data is in reasonable agreement with the theory of Schklovskii~\cite{shklovskii_1976}, which predicts the nonlinearity in the form $I\propto\exp{[eV_{sd}/k_BT\cdot L_C/L]^{1/(1+\nu)}}$, where $L_C$ is the correlation length and $\nu=4/3$ is the 2D critical index~\cite{schklovskii_efros}. Linear dependencies of the form  $\log I\sim [eV_{sd}/k_BT]^{3/7}$ are demonstrated in fig.~\ref{fig4}. This data is taken at $T=0.47$~K for linear response resistances in the range 0.5~M$\Omega$~$<R_\Box<20$~M$\Omega$. The respective  Mott temperatures vary by a factor of $\sim$7 (cf. fig.~\ref{fig2}b). The theory~\cite{shklovskii_1976} is consistent with the experiment in the range $eV_{sd}\geq2k_BT$, which includes the nonlinearity threshold $V_{th}$ defined above, see fig.~\ref{fig1}b.  Surprisingly, the slopes of the linear dependencies are almost independent of $R_\Box$ and correspond to a correlation length of $L_C\approx11$~$\mu$m (see the dashed line fig.~\ref{fig4}). This value of $L_C$  exceeds the sample size and gives an estimate of the localization radius $a>130$~nm. Alternatively, one can evaluate $a$ from the exponential magnetoresistance in perpendicular magnetic fields~\cite{schklovskii_efros}. Fitting such a measurement to a theory~\cite{raikh_exp} we get a different estimate of $a\sim50$~nm for $R_\Box\sim5$~M$\Omega$. Such a large discrepancy between the two methods might be related to the fact that the calculations~\cite{shklovskii_1976,raikh_exp} are performed for the semiconductor impurity band, whereas in our samples the localization occurs in a smooth disorder potential.

As demonstrated in figs.~\ref{fig3} and~\ref{fig2}, our samples exhibit the Poissonian shot noise and behave like macroscopic VRH insulators. Below we argue that this effect is a result of a finite-size effect in the VRH conduction~\cite{rodin,xu}. The key properties of the VRH conduction are captured by a model of the Miller-Abrahams random resistor network (see, e.g.,~\cite{schklovskii_efros}). The model assumes that a pair of localized states, separated by a distance $r_{ij}$,  is connected by a resistor $R_{ij}=R_0\exp{\xi_{ij}}$, where $\xi_{ij}=2r_{ij}/a+\varepsilon_{ij}/k_BT$ and $\varepsilon_{ij}$ is determined by the energies of the two states and the chemical potential. The percolation theory calculates a resistivity of this network  by connecting only those resistors with $\xi_{ij}\leq\xi_C$, where $\xi_C=(T_0/T)^{1/3}$ is a percolation threshold~\cite{schklovskii_efros}. This applies for localized states separated by an average distance (hop length) $l_T=a\xi_C$. The resistivity is determined by the hard hops with the resistances $R_0\exp{\xi_C}$ and the correlation length $L_C=l_T(T_0/T)^{\nu/3}=a(T_0/T)^{(\nu+1)/3}$, where $\nu=4/3$ is the 2D critical index.

In our samples $L_C$ is varied by changing the $T$ or the gate voltage. Although the experimental uncertainty in $a$ does not permit a reliable estimate of the correlation length, it seems natural that $L_C$ decreases at increasing $T$ and for smaller sample depletions. As seen from fig.~\ref{fig3}, this corresponds
to the decrease of the Fano factor from $F\approx1$ to sub-Poissonian values because of the noise averaging out in long samples~\cite{korotkov_PRB2000}. The dependence of $F$ on the ratio $L/L_C$  can be evaluated from fig.~\ref{fig3}b, since $L_C\propto T^{-7/9}$. The observed change in $F$ is roughly twice smaller than follows from the asymptotic law $F\propto(L_C/L)^{\sim0.8}$ at $L\gg L_C$~\cite{sverdlov,camino}, which is qualitatively consistent with the saturation of $F$ at $L\sim L_C$ in numerical calculations~\cite{korotkov_JPhysCM1999}. Hence the onset of the Poissonian shot noise at low-$T$ and strong depletions in our samples is a result of the finite-size effect $L\sim L_C$ in the VRH regime.

Transport measurements give further evidence for the finite-size effect in the VRH regime. In a finite sample, the percolation threshold falls below that for an infinite system~\cite{rodin}. This gives rise to a low $T$ deviation of the dependencies $G_\Box(T)$ from the Mott law, as indeed observed in fig.~\ref{fig2}b. For small deviations one has~\cite{rodin} $\ln G\propto-(T_0/T)^{1/3}+0.25(T_P/T)^{7/3}$, where $T_P$ is a crossover temperature at which $L\sim L_C$. The data of fig.~\ref{fig2}b is in reasonable agreement with this formula with $T_P$ as a fit parameter (dashed lines). The $T$ dependence of the $I$-$V$ nonlinearity
is also consistent with the finite-size effect. According to~\cite{shklovskii_1976}, for $L\sim L_C$  the electrochemical potential drops across a single hard hop, which results in a linear $T$-dependence $V_{th}\propto T$ of the threshold voltage where the nonlinearity sets in.  As seen from fig.~\ref{fig1}b, this is indeed the case in our samples (see fig.~\ref{fig1}b). Consistently, the correlation length extracted from the $I$-$V$ curves exceeds the sample length at low-$T$ (fig.~\ref{fig4}).

As briefly discussed in the introduction, the shot noise in long samples is not universal and averages out in the limit $L\gg L_C$~\cite{kuznetsov,camino,korotkov_PRB2000,savchenko_PSS2004}. Below we discuss the shot noise in the most interesting limit $L\sim L_C$. In this regime the Miller-Abrahams network breaks up into quasi-1D filaments~\cite{rodin} connecting the source and drain reservoirs in parallel. Our observation of the Poissonian shot noise in such a system is puzzling, because of a large number of hops involved. For instance the upper curve in fig.~\ref{fig3}a corresponds to at least $L_C/l_T\sim8$ hops per filament. Since their contributions to the total current and noise are additive, in the following we consider the noise of just one such filament. In a circuit model~\cite{korotkov_PRB2000}, the $i$-th hop in a path is treated as an independent Poissonian noise source with a resistance $R_{i}=R_0\exp{\xi_i}$. In the VRH regime $R_i$ belong to a geometrical sequence with a common ratio of $\sim3$~\cite{schklovskii_efros}, which results in $F=\sum{R_i^2}/(\sum{R_i})^2\approx0.5$. In order to obtain $F=0.9$, which is the lowest bound to the experimental value, an unreasonably broad distribution of $R_i$ has to be assumed (a common ratio of 20). This scenario is unlikely, especially in the regime $L\sim L_C$, where the resistance network is more uniform than in an infinite sample~\cite{rodin}.

We propose a classical approach to the shot noise in the VRH regime based on an analogy with the shot noise in the vacuum tube. Consider a chain of $N$ localized states (sites) connected by random hopping rates $\Gamma_i$ among which the smallest rate $\Gamma_H$ corresponds to the hard hop in the middle. In the nonlinear regime, the electrochemical potential difference across the chain is large ($eV\gg k_BT$) and the hopping occurs preferably in one direction, say from the left to the right. In the limit $\Gamma_H\rightarrow0$, the sites on the left (right) hand side from the hard hop are occupied (empty) most of the time. In this situation, the current occurs via rare injection events across the hard hop and consists of a single vacancy hopping on the left and a single electron hopping on the right (inset of fig.~\ref{fig3}b). Obviously, the circuit model of independent resistances cannot be adequate to such a discrete transport. In fact, the process in question is a close analogue of the transport in a vacuum tube: the vacancy/electron is injected with the Poissonian statistics and moves freely towards the source/drain, which gives rise to $F=1$. It is straightforward to understand what happens if we increase $\Gamma_H$. The injection rate increases, unlike the average dwell time in the chain $T_{dwell}=\sum(\Gamma_i)^{-1}$, where the summation is performed over the sites on the left or right hand side from the hard hop, respectively, for the vacancies or the electrons. Hence, the average occupation probabilities are given by $\rho=2\Gamma_HT_{dwell}/N$. In the finite-size regime in our experiment $N\sim10$ and the rates $\Gamma_i\propto\exp{(-\xi_i)}$ are described by the VRH theory~\cite{schklovskii_efros}, so that $\rho\sim0.1$. Small occupation probabilities indicate independent motion of the electrons/vacancies, therefore the above analogy with the vacuum tube should hold and one expects the Poissonian shot noise.

The vacuum tube analogy can be supported by a purely classical model of Ref.~\cite{derrida}, which has been used for modeling the shot noise in the VRH regime~\cite{korotkov_PRB2000}. Refs.~\cite{derrida} consider a so-called open boundary asymmetric exclusion process (ASEP), which describes a hopping of particles on a uniform 1D lattice. The hopping is allowed in one direction, unless the neighboring site is occupied. Remarkably, the current fluctuations in this model do not necessary average out in an arbitrary long lattice, unlike in calculations of Refs.~\cite{sverdlov,korotkov_JPhysCM1999}. Moreover, in the regimes of low/high density the exact solution for the Fano factor reads $F=1-2\rho$, where $\rho<1/2$ is, respectively, the occupation probability of particles/vacancies~\cite{derrida}. In both cases, the sub-Poissonian noise suppression is a result of negative on-site correlations. If, however, $\rho\rightarrow0$ the correlations are unimportant and the noise becomes Poissonian, just like in the original Schottky's problem~\cite{korotkov_PRB2000}. Formally, our estimate of $\rho\sim0.1$ in the VRH regime corresponds to $F\approx0.8$, much  closer to the experiment. This indicates that the classical approach is a promising framework to understand the shot noise in the VRH conduction.

In summary, we investigated the shot noise in a macroscopic VRH insulator based on a 2D electron system in GaAs. At low $T$ and strong enough depletion
the shot noise close to the full Poissonian value ($F=1\pm0.1$) is observed, which is interpreted as a manifestation of the finite-size effect. We propose a simplified classical approach capable to explain this result and apparently consistent with the VRH conduction theory. Our results call for revision of shot noise theory in the VRH conduction and open up a possibility for accurate quasiparticle charge measurements in nontrivial many-body insulating states~\cite{monceau,shahar,baturina,5/2}.

We gratefully acknowledge the discussions with K.E.~Nagaev, Yu.V.~Nazarov, A.A.~Shashkin, E.V.~Deviatov and V.T.~Dolgopolov. The work was supported by the Russian Ministry of Sciences, RAS, and RFBR under grants Nr.~12-02-00573-a, Nr.~12-02-31404 and Nr.~13-02-00095.


\begin{thebibliography}{50}
\bibitem{remarkPoisson} The Fano factor in our devices stays close to unity at strong depletions. A slightly superpoissonain noise $F\approx1.15$ was observed at the lowest $V_g$, accompanied by a $\sim10$\% asymmetry of the dependencies $S_I(I)$, as, e.g., for the topmost curve in fig.~\ref{fig3}a.
\end{thebibliography}
\end{document}


\maketitle

\begin{figure}[t]
 \begin{center}
  \includegraphics[width=0.6\columnwidth]{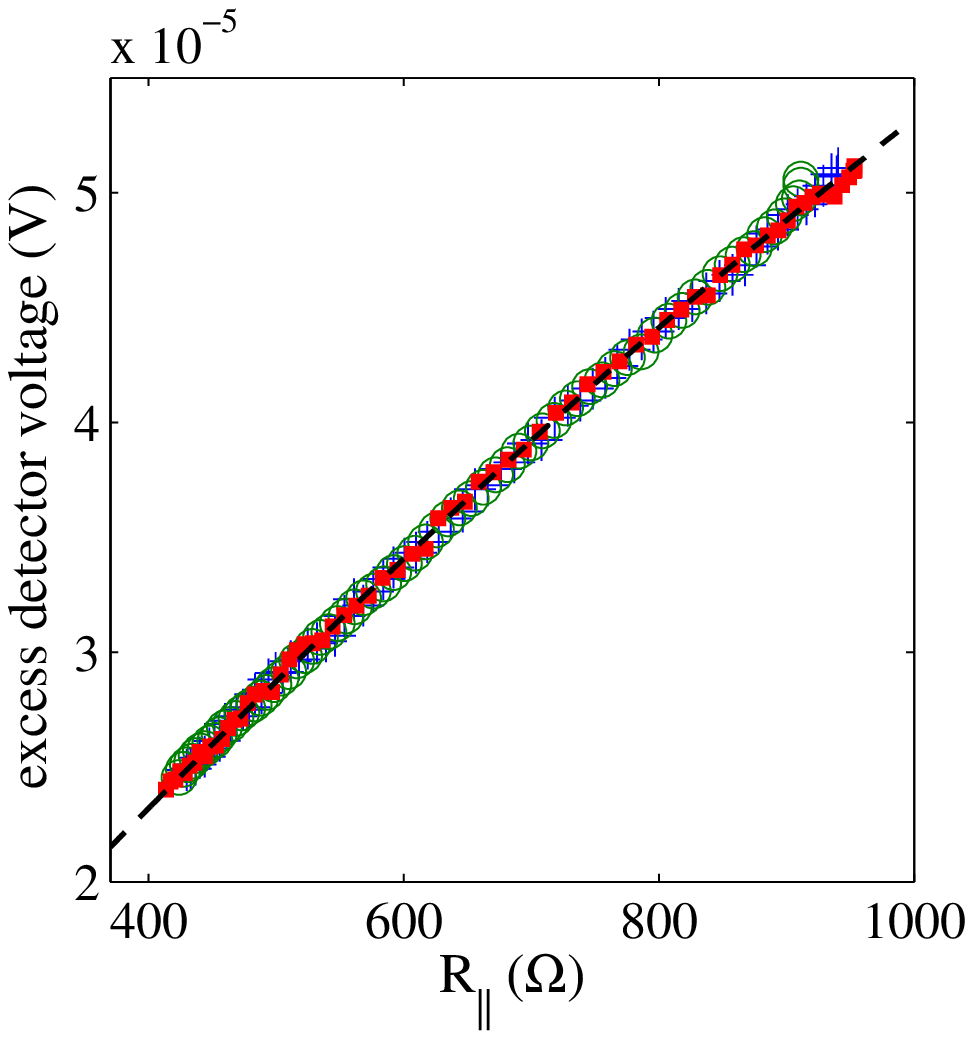}
   \end{center}
  \caption{Fig.~\ref{fig_nyquist}.~Measurement of the equilibrium Johnson-Nyquist voltage noise in the dilution refrigerator. The excess voltage on the detector (normalized by $T$) as a function of the resistance of the sample and load $R_0$ in parallel. Different symbols correspond to the different bath temperatures: circle -- 1.23~K, plus -- 0.69~K, square -- 0.125~K. The dashed line is the fit corresponding to the Johnson-Nyquist noise shunted by a stray capacitance of 3.5~pF. Note that the bath $T$ measured with a thermometer next to the sample didn't go below $\approx120$~mK because of the power dissipated by the low-$T$ amplifier.}\label{fig_nyquist}
\end{figure}